# Velocity-field characteristics and device performance in nanoscale amorphous oxide Thin-Film-Transistors

Chankeun Yoon[1,2] Xiao Wang,[1,2] Jatin Vikram Singh[1,2], Sanjay K. Banerjee[1,2], and Ananth Dodabalapur[1,2,*]

[1] *Chandra Family Department of Electrical and Computer Engineering, The University of Texas at Austin, Austin, Texas 78712, USA*

[2]*Microelectronics Research Center, The University of Texas at Austin, Austin, Texas 78758, USA*

**Email: ananth.dodabalapur@engr.utexas.edu**

**Abstract**

The electron velocity-electric field characteristics in short channel length (50-100 nm) amorphous oxide field-effect transistors (FETs) are described using measured experimental data from indium gallium zinc oxide (IGZO) FETs in conjunction with a physics-based model. Such understanding is crucial for the design of FETs for emerging applications such as in back-end-of-line circuitry for advanced memories and artificial intelligence hardware. In such semiconductor systems, there is an interplay between trapping and extended state (band) transport that has to be considered in detail for a more complete physical understanding of device operation. The approach described in this paper demonstrates such a method and its use for an exemplary semiconductor IGZO. It can be used in many emerging thin-film semiconductors, including several amorphous oxide semiconductors. The carrier mobility is calculated for dominant scattering mechanisms such as trapped carrier scattering and optical phonon scattering. The carrier velocity is computed from the mobility using a modified Caughey-Thomas equation. The physical model considers contact resistance, Joule heating, and electric-field-induced carrier heating, all of which are very important in small geometry FETs. The carrier velocity exhibits a tendency to saturate at high electric fields and reaches values $> 2\times10^6$ cm/s when averaged over all induced carriers (both trapped and in the band) and $> 4\times10^6$ cm/s for carriers in the band.

**Introduction**

As field-effect transistors (FETs) are scaled down in size in any semiconductor technology, it becomes important to get a detailed understanding of high-field charge transport, carrier velocity and its dependence on electric field, and the electrical characteristics of scaled down devices more generally. These are well understood for FETs based on established semiconductor technologies such as silicon and III-V semiconductors (1-3), but such understanding is lacking in FETs based on many emerging thin-film semiconductors such as amorphous oxide semiconductors. For amorphous oxide semiconductors such as indium gallium zinc oxide (IGZO) (4-11) zinc tin oxide (12,13), indium oxide (14,15), etc., very little is known about charge transport in nanoscale devices and at high electric field ($10^4$ V/cm and higher). Earlier work by our group showed that the carrier velocity in IGZO tends to saturate, but this behavior is strongly influenced by traps (16). This earlier work considered IGZO thin-film transistors (TFTs) with channel lengths in the micron scale, often employing thick gate dielectrics and with operating voltages > 10 V.

In this paper, carrier velocity and high-field charge transport are examined both theoretically and experimentally using small channel-length TFTs (channel length $L$ = 50 nm and 100 nm) designed for back-end-of-line applications such as memories and artificial intelligence hardware (17-21). The electrical characteristics (output characteristics) are examined and shown to depend on a multitude of parameters including device dimensions, material properties (particularly trap density and trap energy distribution), thermal properties of component layers, and contact resistance. A detailed model that describes the experimental data very well is presented and is used to study the carrier velocity-electric field characteristics.

A physics-based charge transport model is first briefly described and is used to fit the output characteristics of sub 100 nm channel length amorphous oxide TFTs. From such analysis, the behavior of internal electric field, charge density, and carrier velocity etc. are extracted as a function of position along the channel. This data and the subsequent analysis will be very useful in designing systems based on such transistors for which the non-quasi-static or dynamic response of the transistor is important. While data is presented for amorphous IGZO, the model and approach are more general and can be applied to a semiconductor system in which trapping plays a significant role.

**Physical Model**

The physical model employed in this work is a quasi-two-dimensional analytical framework based on multiple trap and release (MTR) charge transport. The basic model is described in detail in (22) and involves self-consistent solution of Poisson's equation and the current continuity equation along the channel. Carrier mobility is calculated for various scattering mechanisms including trapped carrier scattering and polar optical phonon scattering and combined using Mattheissen's rule. These two scattering mechanisms dominate in high-mobility amorphous oxides such as IGZO and zinc tin oxide (16,22,23). Other scattering mechanisms such as interface roughness scattering and surface phonon scattering have been analyzed in detail by our group in previous work, and do not impact the mobility significantly in high-mobility FETs such as those employed in this study. The most important feature of this model is that the details of trap states (concentration and energy distribution) are considered together with band transport influenced by multiple scattering mechanisms. This model is also described in detail in **Supporting Information 1** and **2**.

In conventional semiconductors such as crystalline silicon and two-dimensional semiconductors such as $MoS_2$ and Black Phosphorus, charge transport is commonly described using band transport based velocity models such as Caughey-Thomas formula (24-26). The carrier velocity $v$ is typically expressed as:

$$v = \frac{\mu_{eff}E}{[1+(\mu_{eff}E/v_{sat})^\beta]^{\frac{1}{\beta}}}, \qquad (1)$$

where $\mu_{eff}$ is the effective mobility, $v_{sat}$ is saturation velocity, $E$ is electric field and $\beta$ is an empirical fitting parameter.

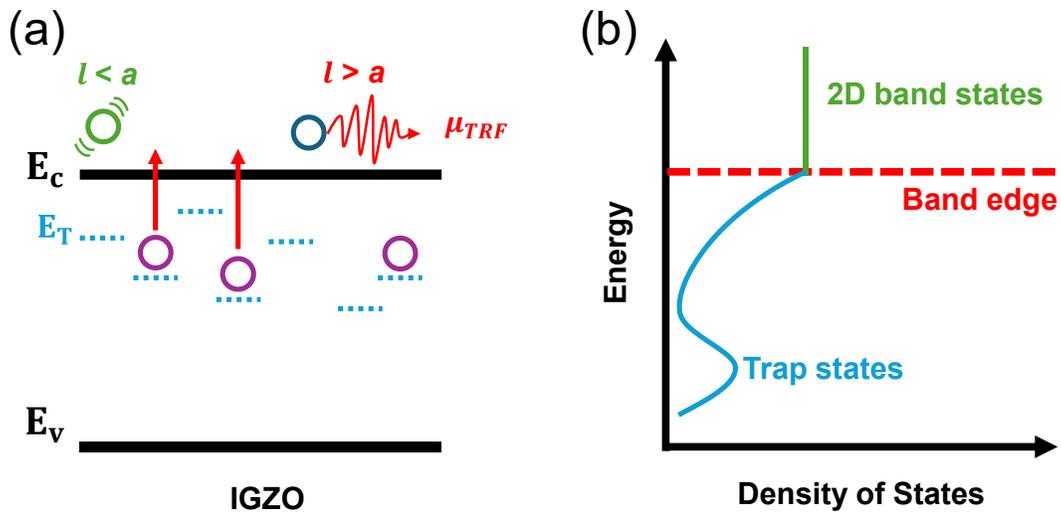

**Figure 1**. Schematic illustration of charge transport in amorphous IGZO based on multiple trap and release (MTR) mechanism. Carriers are excited from localized trap states into extended states. Only carriers with a path length ($l$) larger than lattice constant ($a$) contribute effectively to conduction. (b) Density of states (DOS) distribution in amorphous IGZO. Localized trap states extend into the bandgap below the band edge, while extended states above the band edge contribute to charge transport.

In contrast, charge transport in amorphous oxide semiconductors such as IGZO is more complex due to disordered nature of the amorphous semiconductor film and the interplay between trapping and band transport. Carriers are distributed between extended states above the conduction band edge and localized tail states within the bandgap. This distribution is influenced both by lattice temperature (affected by Joule heating in small geometry devices) and electric-field induced carrier heating. Therefore, transport is better described within the framework of an extended multiple trap and release (MTR) model (16,22). Only a subset of carriers with sufficiently long path length (> lattice constant, *a*) contribute effectively to conduction, while the remaining carriers are dynamically localized by strong scattering (27) as illustrated in **Fig.1a**. This basic approach has been used by our group in studying the effects of various scattering mechanisms such as trapped carrier scattering (16,22), interface roughness scattering (28), and various phonon scattering mechanisms on low-field mobility in amorphous oxide semiconductors and organic semiconductors (29), with which very small geometry FETs have been reported (30).

The carrier distribution in an amorphous oxide semiconductor is schematically illustrated in **Fig. 1b**. The electronic density of states is assumed to be two-dimensional above the band edge and possesses a high concentration of tail states that extend into the bandgap (13). The density of states within the bandgap is assumed to be exponential (16,22). At any temperature, carriers are thermally distributed between the trap states in the bandgap and extended states above the band edge. At high electric fields, the field can provide additional energy and the carrier distribution is characterized by an effective carrier temperature, $T_{eff}$, that can be higher than the lattice temperature (22):

$$T_{eff} = \sqrt{T^2 + (YeaE/k_B)^2}, \qquad (2)$$

where *E* is lateral electric field in the channel, *a* is lattice constant, $k_B$ is Boltzmann constant, and *Y* is assumed to be 1. Here, *T* denotes the local lattice (channel) temperature which will be described later.

To properly describe carrier velocity and its variation with electric field in such disordered semiconductors, a modified Caughey-Thomas velocity (*v*)-electric field (*E*) relationship is used for amorphous oxide semiconductors (16), given by:

$$v = \frac{P_{MTR} P_{TRF} \mu_{TRF} E}{[1+(\mu_{TRF} E/v_{sat})^\beta]^{\frac{1}{\beta}}}, \qquad (3)$$

where $P_{MTR}$ represents the ratio of carriers in extended states to the total induced carriers, $P_{TRF}$ is a phenomenological transport reduction factor accounting for carriers with sufficiently long path length (31), $\mu_{TRF}$ is mobility of carriers considering phenomenological transport reduction factor, $E$ is electric field in the channel, $v_{sat}$ is saturation velocity of trap-free IGZO, and $\beta$ is a fitting parameter. The product $P_{MTR} P_{TRF} \mu_{TRF}$ is equivalent to the experimentally measured low-field mobility in FETs and is illustrated in **Supporting Information 3 (Fig. S2)**. It is noted that in amorphous semiconductors such as IGZO, with high mobility, evidence of crystalline order on the scale of a several nm has been reported (32-34). This is larger than the mean free path, which is in the range 1-3 nm. This justifies the treatment of this semiconductor as a crystal for purposes of charge transport analysis.

The velocity given by **Equation 3** can be described as the effective carrier velocity in that it represents an average velocity for *all* gate-induced carriers, including many that are trapped. In most of the following discussion, the effective carrier velocity or effective velocity is considered. The band velocity, on the other hand, is given by **Eq. 3**, but without the $P_{MTR}$ term. The band velocity considers only carriers that in extended states above the band-edge and includes a fraction that are dynamically trapped (by strong scattering). The band velocity is also described in the following sections.

**Equation 3** can be used to describe carrier velocity in amorphous oxide semiconductors. The product $P_{MTR} P_{TRF}$ can be understood to be the fraction of gate-induced charge carriers that participate in band transport. $P_{MTR}$ is simply the fraction of gate-induced carriers that occupy extended states. The closer this product approaches 1, the better the electronic quality of the semiconductor. In trap-free semiconductors, $P_{MTR} = 1$. In semiconductors (Ex. Silicon) with high mean free path, the transport reduction factor, $P_{TRF} = 1$. In amorphous oxide semiconductors, which have relatively small mean free paths (in comparison with silicon), it becomes necessary to consider that there is a distribution of path lengths and to allow for some carriers in extended states, but possessing smaller path lengths, to be dynamically localized. The factor $P_{TRF}$ accounts for such dynamically localized carriers. Thus $P_{MTR}P_{TRF}$ can be understood to be the fraction of gate-induced charge carriers that actually participate in band transport. The closer this product approaches 1, the better the electronic quality of the semiconductor.

In order to accurately evaluate charge transport, the voltage drops across the source and drain contacts must be subtracted from the applied source-drain and gate-source voltages. An initial estimate of the contact resistance at low drain voltage is obtained using transfer-length method (TLM) analysis across multiple channel lengths and gate voltages. This TLM-based contact resistance serves as a baseline and is subsequently refined into a $V_{GS}$ and $V_{DS}$-dependent $R_c$ model within the self-consistent transport framework, as described in **Supporting Information 1 and 2**. As mentioned above, both Joule heating and carrier heating caused by high electric fields have to be considered as these can be significant in small geometry FETs with high current densities (25, 35). The thermal resistance ($R_{TH}$) is modeled using a simplified vertical heat-flow approximation as:

$$R_{TH} \approx R_{channel} + R_{TBR} + R_{oxide} + R_{contact}, \qquad (4)$$

where $R_{channel}$ is thermal resistance of IGZO channel, $R_{TBR}$ is thermal boundary resistance at the IGZO–Al$_2$O$_3$ and Al$_2$O$_3$–Ni interfaces, $R_{oxide}$ is thermal resistance of the Al$_2$O$_3$ dielectric layer and $R_{contact}$ is thermal resistance of Ni contact. A vertical thermal resistance ($R_{TH}$) of 9×10$^4$ K/W is used in this work and the corresponding thermal parameters are summarized in **Supporting Information 2**. The channel temperature due to Joule heating is expressed as:

$$T = T_0 + R_{TH}I_D V_{channel}, \tag{5}$$

where $T_0$ is the ambient temperature and $V_{channel}$ is the total potential drop in the channel from source to drain.

The method outlined above, is described in more detail in **Supporting Information 1 and 2** and enables the simultaneous consideration of the important physical effects that affect carrier velocity in a disordered semiconductor FET. Importantly, it explicitly considers the main scattering mechanisms that determine low-field mobility including trapped carrier scattering and the dominant phonon scattering mechanism which is optical phonon scattering.

## Results and Discussion

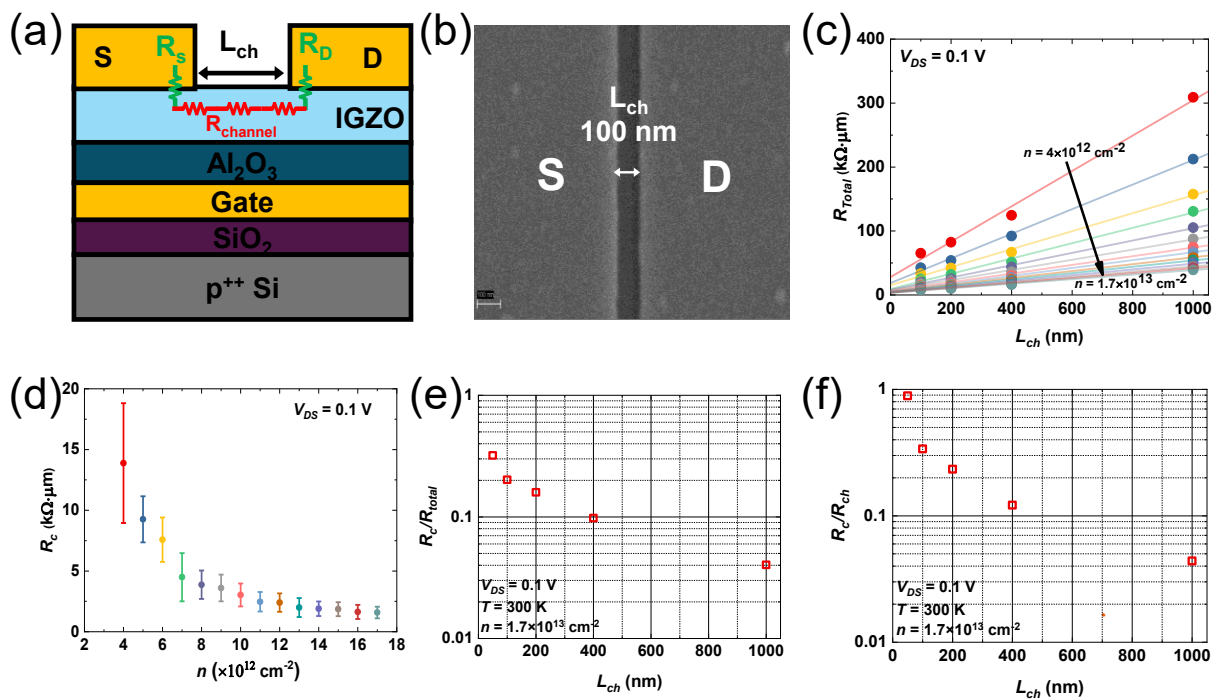

**Figure 2**. (a) Schematic illustration of bottom-gate, top-contact IGZO thin-film-transistor, showing major resistance components: source contact resistance ($R_s$), drain contact resistance ($R_D$) and channel resistance ($R_{channel}$). (b) Top-view scanning electron microscopy (SEM) image of 100 nm $L_{ch}$ device. (scale bar: 100 nm) (c) Total resistance ($R_{total}$) as function of channel length measured at $V_{DS}$ = 0.1 V for different carrier densities. Symbols represent experimental data and solid lines are linear fits. (d) Extracted $R_c$ obtained from the y-intercepts of the linear fits to $R_{total}$ vs. $L_{ch}$ in **Fig. 2c**. Error bars represent the standard errors of the fitted intercepts. (e) $R_c/R_{total}$ and (f) $R_c/R_{channel}$ as a function of $L_{ch}$ at $n = 1.7 \times 10^{13}$ cm$^{-2}$ and $V_D$ = 0.1 V. For (e) and (f), data from 50 nm $L_{ch}$ device are also included, and additional data for the 50 nm device are provided in **Supporting Information**.

**Figure 2a** shows the device structure of the bottom-gate, top-contact IGZO thin-film-transistor used in this study, with channel lengths ($L_{ch}$) ranging from 50 nm to 1000 nm. The total device resistance can be decomposed into three primary components: source contact resistance ($R_S$), drain contact resistance ($R_D$) and channel resistance ($R_{ch}$). Therefore, the externally applied $V_{DS}$ and $V_{GS}$ can be expressed as:

$$V_{DS} = V_{DS(channel)} + I_D \times (R_S + R_D), \quad V_{GS} = V_{GS(channel)} + I_D \times R_S$$

where $V_{DS(channel)}$ and $V_{GS(channel)}$ are used in modeling charge transport within the semiconductor channel. **Figure 2b** shows a top-view scanning electron microscope (SEM) image of a 100 nm $L_{ch}$ device. Details of the device fabrication process are provided in **Methods** section. To estimate contact resistance ($R_c$), the total device resistance ($R_{total}$) is plotted as a function of $L_{ch}$ at $V_{DS} = 0.1$ V for different carrier densities ($n \approx C_{ox}V_{ov}/q$), as shown in **Fig. 2c**. From a linear fit of $R_{total}$ vs $L_{ch}$, $R_c$ is extracted from the y-intercept, assuming symmetric source and drain contacts and contact resistance (**Fig. 2d**). **Figure 2e** and **2f** shows the ratio $R_c/R_{total}$ and $R_c/R_{ch}$ as a function of $L_{ch}$ at $n \sim 1.7 \times 10^{13}$ cm$^{-2}$. As $L_{ch}$ decreases, both $R_c/R_{total}$ and $R_c/R_{ch}$ increase, indicating that $R_c$ becomes increasingly dominant especially in the sub-100-nm $L_{ch}$ regime. These results highlight the importance of considering contact resistance when analyzing high-field carrier transport and velocity of nanoscale oxide TFTs. The TLM method used to estimate contact resistance is also used by several researchers in the field, even though it may be less accurate at small channel lengths (36-38). It is seen in **Fig. 2e** and **2f** that contact resistance effects can become dominant at small channel lengths.

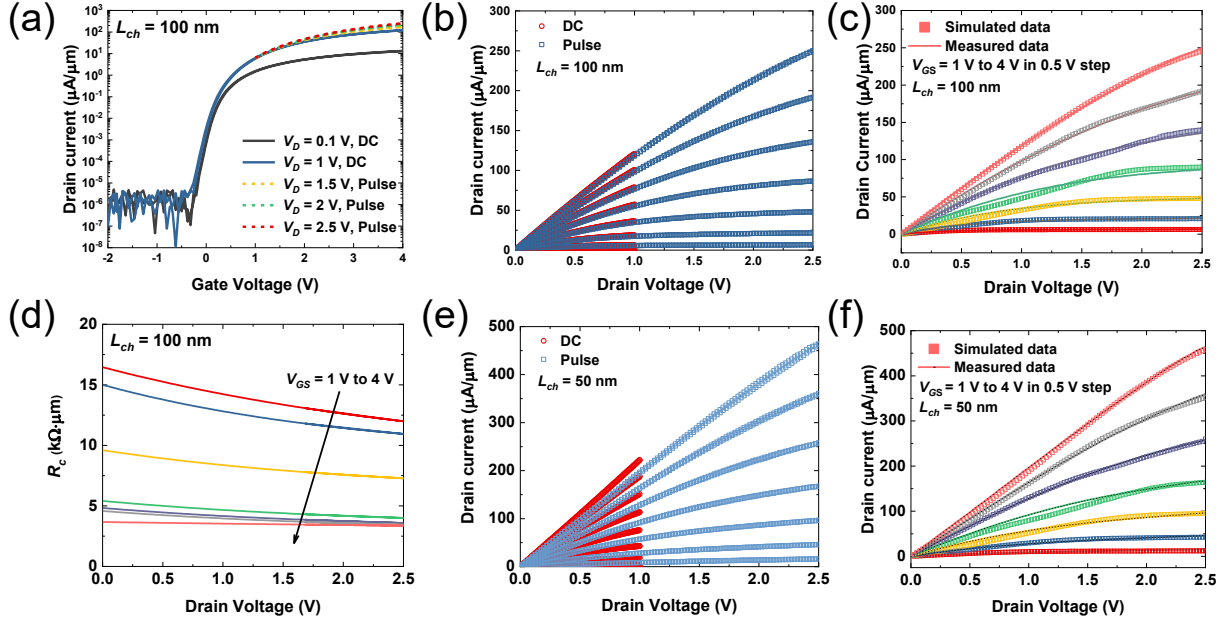

**Figure 3**. (a) $I_D$–$V_G$ curve of 100 nm $L_{ch}$ device. DC measurements are performed at $V_D = 0.1$ V and 1 V, while pulse measurements are used at higher $V_D$ ($V_D = 1.5$ V, 2 V and 2.5 V). For pulse measurements, $V_G$ was swept from 1 V to 4 V. (b) $I_D$–$V_D$ curve of 100 nm $L_{ch}$ device. DC measurements are performed at $V_D = 0 \sim 1$ V and $V_G = -2$ V to 4 V. Pulse measurements are performed at $V_D = 0 \sim 2.5$ V and $V_G = 1$ V $\sim$ 4 V. (c) $I_D$ – $V_D$ curve of the 100 nm $L_{ch}$ device, showing experimental data and simulated results obtained from the self-consistent quasi-2D Poisson-continuity model. (d) $V_G$ and $V_D$ dependent $R_c$ used in the model. (e) $I_D$–$V_D$ curve of 50 nm $L_{ch}$ device under the same bias conditions as in (b). (f) $I_D$–$V_D$ curve of the 50 nm $L_{ch}$ device, showing experimental data and simulated results obtained from the self-consistent quasi-2D Poisson-continuity model

**Figure 3a and 3b** shows measured DC transfer and output characteristics of a $L_{ch}$ =100 nm device. Initial DC measurements are performed at $V_D = 0.1$ V and 1 V. To further increase $V_D$ while mitigating self-heating effects, pulsed measurements are used for $V_D$ up to 2.5 V and compared with DC measurement (**Fig. 3b**). The close agreement between DC and pulsed data may be noted for $V_D$ values up to 2.5 V. Following the electrical measurements, the device characteristics are simulated using a self-consistent quasi-2D Poisson-continuity framework. **Figure 3c** compares the experimentally measured and simulated output characteristics for a 100 nm channel length FET. The model reproduces the experimental data with good agreement. The model includes Joule (lattice) heating as well as carrier heating by the electric field. **Figure 3d** shows the extracted $V_G$ and $V_D$ dependent contact resistance $R_c(V_G, V_D)$ used in the model. Compared to $R_c$ values extracted using the TLM in **Fig. 2d**, the $R_c$ values used in model are slightly higher. This difference likely arises because conventional TLM analysis assumes linear scaling of $R_{total}$ down to $L_{ch} = 0$, which may not remain valid in the sub-100 nm channel length regime. As $L_{ch}$ decreases, $R_c$ tends to decrease sublinearly, leading to underestimation of the $R_c$ as illustrated in **Fig. S3**. As illustrated in **Fig**. **3d**, $R_c$ decreases with increasing $V_G$ due to increased carrier accumulation near the source contact, which reduces the effective Schottky barrier height and improves carrier injection into the channel.

Data from an $L_{ch} = 50$ nm channel length FET are shown in **Fig. 3e**, in which the DC measurement data and pulsed measurement data for output characteristics are overlaid. Simulated data for $L_{ch} = 50$ nm are overlaid with pulsed experimental data in **Fig. 3f**, again showing good agreement and providing additional validation of the modeling approach employed. Additional data used in 50 nm device simulation are illustrated in **Fig.S4**.

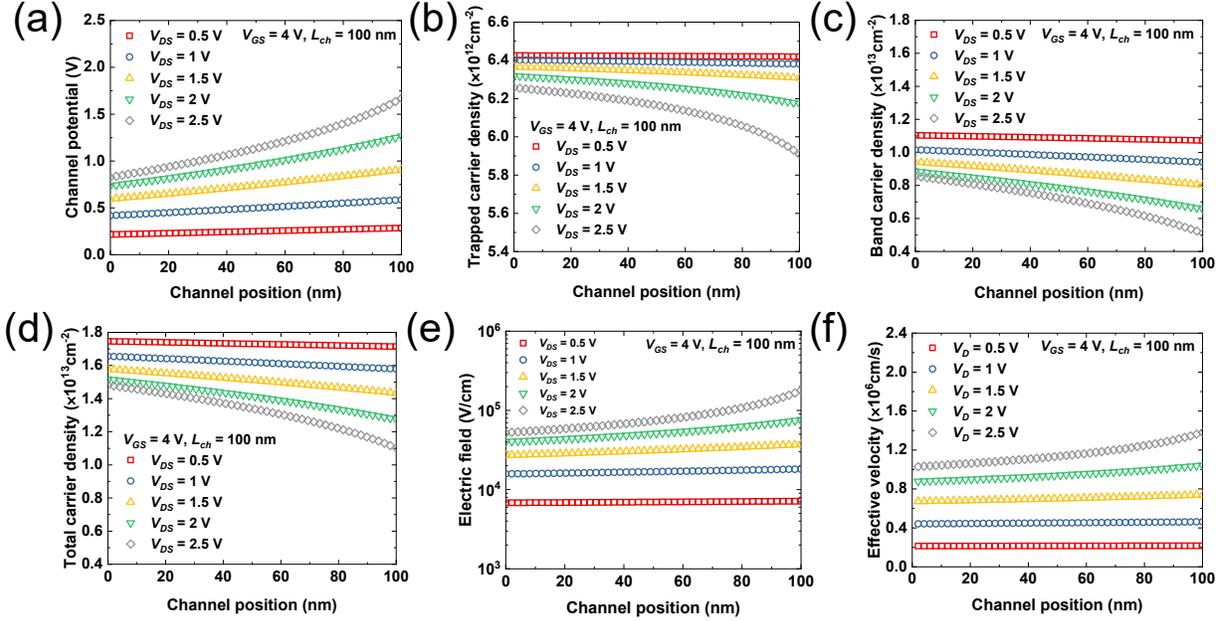

**Figure 4**. Extracted spatial profiles along the channel of a 100 nm $L_{ch}$ device at $V_{GS}$ = 4 V: (a) channel potential $V_{ch}(y)$, (b) trapped carrier density $n_{trap}(y)$, (c) band carrier density $n_{band}(y)$, (d) total carrier density $n_{total}(y) = n_{band}(y) + n_{trap}(y)$, (e) lateral local electric field $E(y)$ and (f) effective carrier velocity $v(y)$. The channel position $y$ = 0 nm and 100 nm correspond to the source and drain side, respectively.

**Figures 4** summarizes spatial profiles along the channel from source to drain ($y$ direction) of the calculated lateral channel potential $V_{ch}(y)$, carrier density $n(y)$, electric field $E(y)$, and effective carrier velocity $v(y)$ for the 100 nm $L_{ch}$ device at $V_{GS}$ = 4 V. Data in **Fig. 4** are results of modeling of experimental data described above. As a consequence of source and drain contact resistances ($R_S$ and $R_D$), a portion of externally applied voltage drops across the contacts. As a result, the internal channel potential near the source side is higher than 0 V, while the potential near the drain side is lower than applied $V_D$ as shown in **Fig. 4a**. In all subsequent modeling data, these potential drops are subtracted from the applied source-drain and gate-source voltages.

The channel potential increases monotonically with both increasing $V_{DS}$ and with channel position toward the drain. The quasi 1-dimensional distribution, along the length of the channel, of trapped carrier density, band carrier density, and total carrier density are shown in **Figs. 4(b)-4(d)**. As expected, the carrier density decreases as the channel position moves toward drain side because the local channel potential, $V_{ch}(y)$ increases which reduces effective gate overdrive voltage. Consequently, higher $V_D$ results in reduced band, trapped and total carrier densities. The lateral local electric field, extracted from the relation $E(y) = dV_{ch}(y)/dy$, is shown in **Fig. 4e**. At low $V_D$ (i.e., 0.5 ~ 1.5 V), the electric field remains fairly uniform along the channel. As $V_D$ increases and the device moves towards the saturation region of operation, the electric field becomes non-uniform and increases non-linearly near the drain. The effective carrier velocity profiles, shown **in Fig. 4f,** are calculated using the equation $v(y) = I_d/(qWn_{total}(y))$ with the drain current being constant along the length of the channel. The effective carrier velocity increases toward the drain side due to enhanced local electric field. Data from 50 nm channel length devices show qualitatively similar behavior and are shown in **Fig. S5**.

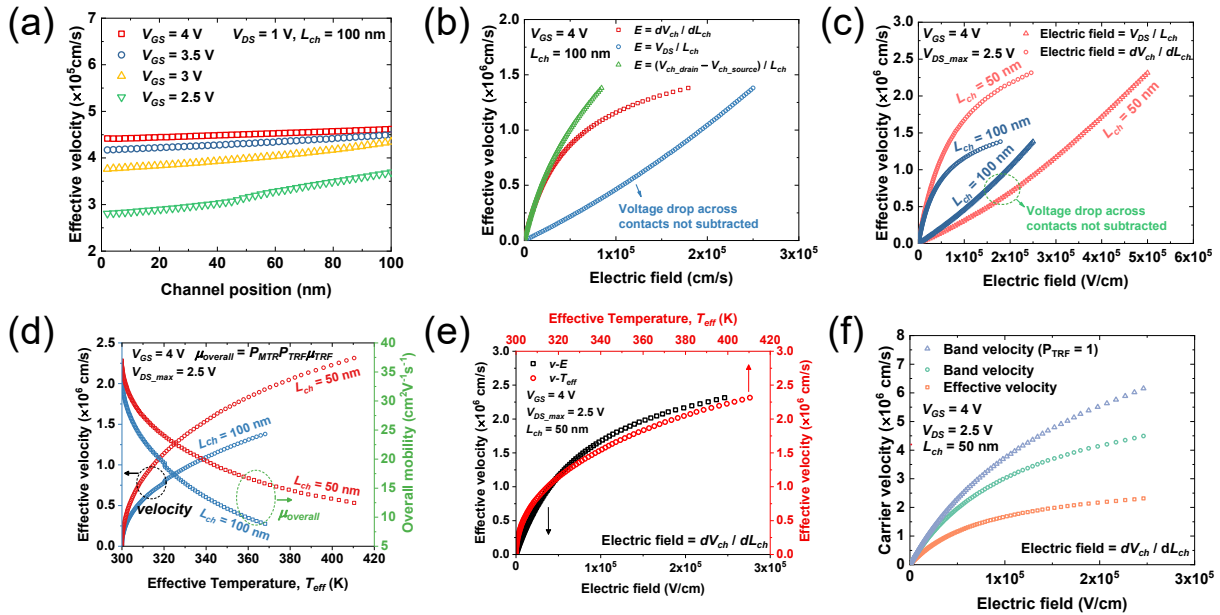

Figure 5. (a) Effective carrier velocity (for all induced carriers) profiles along the channel at $V_{DS}$ = 1 V for different $V_{GS}$. (b) Effective carrier velocity as a function of electric field at drain side for $V_{GS}$ = 4 V. The electric field is defined either using applied drain voltage ($E=V_{DS}/L_{ch}$) or internal channel voltage excluding contact voltage drop ($E=(V_{ch,drain}-V_{ch,source})/L_{ch}$) or local electric field ($E=dV_{ch}/dL_{ch}$). (c) Comparison of effective carrier velocity vs. electric field for $L_{ch}$ = 50 nm and 100 nm, highlighting impact of voltage drop across contacts. (d) Effective carrier velocity and thin

film mobility as a function of local electric field for $L_{ch}$ = 50 nm and 100 nm. (e) Effective carrier velocity as a function of effective temperature and local electric field for $L_{ch}$ = 50 nm. (f) Band velocity as a function of local electric field. The band velocity is calculated from the effective carrier velocity in conjunction with occupancy statistics. The highest velocities are attained if the transport reduction factor ($P_{TRF}$) = 1, which means that all carriers in the band have a path length greater than the lattice spacing for this semiconductor.

To evaluate gate voltage dependence of effective carrier velocity, the spatial profiles of effective carrier velocity along the channel are plotted as function of $V_{GS}$ at $V_{DS}$ = 1 V, as shown in **Fig. 5a**. At this moderate drain bias, the effective carrier velocity increases with increasing $V_{GS}$ over the entire channel. This trend contrasts with conventional crystalline Si and two-dimensional semiconductors, where higher $V_{GS}$ often results in reduced effective carrier mobility due to enhanced polar phonon scattering (1). In IGZO, the density of carriers excited to the band with increasing $V_{GS}$ increases the effective velocity (which is the velocity averaged over all field-induced carriers) and mobility. This increase more than compensates for any decrease due to enhanced polar phonon scattering.

**Figure 5b** shows the behavior of effective carrier velocity as a function of electric field. The magnitude of the electric field needs to be computed carefully taking into account the potential drops at the electrodes. The difference in behavior with and without subtracting these contact potential drops is clearly seen in **Fig. 5b**. When electric field is calculated without subtracting the drops at the contacts, the velocity-field characteristic is supra-linear and is dominated by contact resistance behavior. The velocity-field characteristic shows a sub-linear behavior that is expected in semiconductors such as Si with band transport. **Figure 5b** also shows the difference between effective carrier velocity as a function of electric field averaged across the channel length ($E = (V_{ch\_drain} - V_{ch\_source})/L_{ch}$) and as a function of local electric field ($E = dV_{ch}/dL_{ch}$).

The effects of contact resistance (and consequent potential drops) are also seen in **Fig. 5c.** In this figure data are shown for both 50 nm and 100 nm channel length devices with and without excluding the voltage drop across contacts in the electric field calculation. For each channel length, the effective carrier velocity is plotted as a function of average electric field in the channel and also the local electric field. It may be noted that the effective velocity in the 50 nm FET is higher than that in the 100 nm FET for the same local electric field ($dV_{ch}/dL_{ch}$). The effective carrier velocities as a function of local electric field for $L_{ch}$ = 50 nm and $L_{ch}$ = 100 nm FETs are also shown in **Fig. 5d** together with the calculated thin-film mobility for each device. The thin-film mobility is given by the product $P_{MTR}P_{TRF}\mu_{TRF}$ (which is defined in the discussion of Eq. 3 and described in **Supporting Information**). This $\mu_{TRF}$ is calculated from the band mobility ($\mu_0$) in the material considering trapped carrier scattering and phonon scattering combined using Matthiessen's rule as described in **Supporting Information 1** ($\mu_{TRF} = (1 - ln(P_{TRF})\mu_0)$. The product $P_{MTR}P_{TRF}$ describes the fraction of gate-induced carriers that are involved in band transport. It is noted that in the low electric field limit, the thin-film mobility will be equivalent to the experimentally measured mobility from FETs. It is observed that the thin-film mobility decreases with increasing electric field. This is expected behavior and is due to enhanced phonon scattering at higher fields. The difference in thin film mobility (and also effective velocity) in $L_{ch}$ = 50 nm and $L_{ch}$ = 100 nm devices is due to differences in carrier density and current density for a given field. It has been pointed out above that the thin-film mobility in amorphous IGZO FETs is highly dependent on carrier density.

**Figure 5e** again plots the effective carrier velocity versus local electric field for a $L_{ch}$ = 50 nm FET. Also shown in the figure is the effective carrier temperature ($T_{eff}$), which depends on the electric field due to Joule heating and electric-field induced carrier heating (**Equation 2**). It is noted that effective temperature reaches ~ 400 K due to high gate and drain voltages, which results in an increase in current density due to more carriers being excited to the band. The use of high drain (and gate) voltages allows the calculation of velocities over a larger range of electric fields, which is desirable in such a study.

**Figure 5f** contains the final results of this study. The effective carrier velocity (which is averaged over all the gate induced electrons) is plotted as a function of local electric field. Effective carrier velocity values of more than $2\times10^6$ cm/s are achieved at local electric fields $\geq 2\times10^5$ V/cm. The corresponding band carrier velocities are also shown. The band carrier velocity is obtained by dividing the effective carrier velocity by the $P_{MTR}$ factor (or the fraction of induced carriers that are excited to extended band states). It is assumed that carriers in trap states do not contribute to conduction, in accordance with the basic premise of multiple trap and release transport. Also, as discussed before, some of the carriers in band states do not contribute to current flow. This is because of the relatively small path length in amorphous IGZO as illustrated in **Fig. 1a**. Carriers with path lengths below one lattice constant are dynamically localized by strong scattering. If we exclude this fraction (which is obtained from $P_{TRF}$), then we get the band velocity assuming all carriers in the band move under the influence of the electric field. This is also plotted in **Fig. 5f**. It is seen that this velocity reaches value $> 6\times10^6$ cm/s at the highest electric fields considered. This is close to theoretically predicted values of the saturation velocity in IGZO (16,39). (this value will depend, of course, on the exact composition of the alloy). The shape of the velocity-field curves indicates a tendency to saturate at high electric fields. This saturation behavior is influenced by changing carrier density in extended states with electric field.

**Figure 5f** also illustrates the gains that can be made in carrier velocity if the material quality of the IGZO is improved (for example, by new film growth methods and annealing). The effective velocity will become greater and will approach the band velocity in the limit of there being negligible trapping. The value of $P_{TRF}$ will also approach 1, as the path lengths and mean free path will also increase with improving crystalline quality. The analysis approach described in this paper will accurately describe the high-field transport and velocity behavior that is essential for understanding device behavior at small geometries and high frequencies. It is noted that the analysis is rooted in a physics-based model that starts from calculating mobilities due to various scattering mechanisms relevant for this material system. The extensive use of experimental data from measured devices that are used in the model leads to accurate calculations of electron velocity values at high electric fields.

**Conclusion**

The characteristics of small geometry ($L_{ch}$ = 50 nm and $L_{ch}$ =100 nm) IGZO FETs have been investigated in detail using experimental data from devices that are input into a detailed physics-based model that considers both trapping and extended state-transport and explicitly considers dominant charge carrier scattering mechanisms in this material system. The effects of contact resistance are also considered as these can become important in small geometry FETs. An important result of this paper is the derivation of the velocity-field characteristic for amorphous IGZO that includes the effects of trapping and extended states transport, and thermal (heating) effects. The effective velocity of all gate-induced carriers is calculated along with the velocity of carriers in the band as a function of electric field. A tendency for the velocity to saturate at high electric fields is clearly demonstrated. The methods described in this paper can also be used for other semiconductor transistors in which both trapping and extended-state transport need to be considered. This paper is one of the first to address the important problem of high-field transport, carrier velocity, and velocity saturation in disordered semiconductor transistors.

**Methods**

**Device Fabrication**

Heavily doped $p^{++}$ Si substrate with 86 nm $SiO_2$ layer is prepared and cleaned using acetone, methanol and isopropanol. Local bottom gate electrodes are defined by electron beam lithography (JBX-8100FS/E), followed by electron beam evaporation of 25 nm Ni using a PRO Line PVD 200 system. A 9 nm thick $Al_2O_3$ gate dielectric is subsequently deposited by atomic layer deposition at 200 ºC using a Fiji F200 system. The 6 nm thick IGZO active channel layer is deposited by RF sputtering using PVD 75 at a power of 150 W at 5 mTorr with 7 % $O_2$ pressure. The IGZO oxide target has a $Ga_2O_3$:$In_2O_3$:ZnO molar ratio of 1:2:2. Following deposition, the samples are annealed on a hotplate at 350 ºC for 1 hour in ambient air. The IGZO channel regions are patterned using the electron beam lithography and wet-etched using diluted hydrochloric acid (HCl) solution. Source and drain electrodes are subsequently formed using electron beam lithography system and 30 nm thick Ni is deposited by electron beam deposition. Finally, a PMMA A4 layer is spin-coated as a passivation layer and annealed at 180 ºC for 2 minutes. Then, contact windows to the gate, source and drain electrodes are defined by electron beam lithography.

**Device Characterization**

Electrical characterization is carried out at room temperature using a B1500A semiconductor parameter analyzer under ambient conditions. Pulsed measurements are used with a pulse width of 1 ms and pulse period of 5 ms.

**AUTHOR INFORMATION**

**Corresponding Author**

Ananth Dodabalapur − Chandra Family Department of Electrical and Computer Engineering, The University of Texas at Austin, Austin, Texas 78712, USA; Microelectronics Research Center, The University of Texas at Austin, Austin, Texas 78758, United States; Email: ananth.dodabalapur@engr.utexas.edu

**Authors**

Chankeun Yoon − Chandra Family Department of Electrical and Computer Engineering, The University of Texas at Austin, Austin, Texas 78712, USA; Microelectronics Research Center, The University of Texas at Austin, Austin, Texas 78758, USA

Xiao Wang − Chandra Family Department of Electrical and Computer Engineering, The University of Texas at Austin, Austin, Texas 78712, USA; Microelectronics Research Center, The University of Texas at Austin, Austin, Texas 78758, USA

Jatin Vikram Singh − Chandra Family Department of Electrical and Computer Engineering, The University of Texas at Austin, Austin, Texas 78712, USA; Microelectronics Research Center, The University of Texas at Austin, Austin, Texas 78758, USA

Sanjay K. Banerjee − Chandra Family Department of Electrical and Computer Engineering, The University of Texas at Austin, Austin, Texas 78712, USA; Microelectronics Research Center, The University of Texas at Austin, Austin, Texas 78758, USA

**Author Contributions**

C.Y., X.Y and A.D. conceived the experiments, simulation and wrote the manuscript. C.Y. fabricated devices and performed electrical characterization and simulation. X.Y. assisted with simulation. J.V.S. and S.K.B. assisted with pulse measurement. The manuscript was written through contributions of all authors. All authors have given approval to the final version of the manuscript.

**Data availability**

All data are available in the paper and Supplementary Information are available from the corresponding author upon reasonable request.

**ACKNOWLEDGMENT**

This work was supported by Keck foundation under grant #26753419. The work was done at the Texas Nanofabrication Facility supported by the National Science Foundation under Grant NNCI-2025227

Supporting information for

# Velocity-field characteristics and device performance in nanoscale amorphous oxide Thin-Film-Transistors

Chankeun Yoon[1,2] Xiao Wang,[1,2] Jatin Vikram Singh[1,2], Sanjay K. Banerjee[1,2] and Ananth Dodabalapur[1,2,*]

[1] *Chandra Family Department of Electrical and Computer Engineering, The University of Texas at Austin, Austin, Texas 78712, USA*

[2] *Microelectronics Research Center, The University of Texas at Austin, Austin, Texas 78758, USA*

Email: **ananth.dodabalapur@engr.utexas.edu**

**Supporting Information 1**

**Charge transport in a-IGZO TFTs**

Electrical properties of a-IGZO TFTs are simulated based on extended multiple trap and release (MTR) model considering its moderate mobility (~10 cm$^2$V$^{-1}$s$^{-1}$) and neglected hopping transport (1). First, the shallow trap density of states (DOS) in a-IGZO can be described by an exponential tail profile extending from the band edge to the bandgap. (2,3):

$$DOS_{trap}(E_k) = \frac{N_T}{k_B T_{ta}} e^{(E_k - E_c)/k_B T_{ta}} \text{ (for } E_k < E_c\text{)}, \qquad (1)$$

where $N_T$ is total trap density, $T_{ta}$ is characteristic temperature of the trap DOS, $E_k$ is carrier energy, $E_c$ is conduction band edge and $k_B$ is Boltzmann constant. The DOS of extended states can be assumed to be two dimensional, since the charge transport in a-IGZO TFTs primarily occurs along the interface between the gate dielectric and semiconductor. The DOS of extended states is expressed as (2,3):

$$\boldsymbol{DOS_{band}(E_k)} = \frac{gm^*}{2\pi\hbar^2} \text{ (for } E_k > E_c\text{)} \qquad (2)$$

where $g$ is the spin degeneracy factor (=2), $m^*$ is the effective mass of a-IGZO and $\hbar^2$ is the reduced Planck constant. The valley degeneracy factor is assumed to be 1.

Based on MTR theory, when a gate voltage is applied, a portion of the carriers is captured by trap sites, while the remaining carriers occupy extended states and contribute to band transport. The equilibrium trapped carrier density ($n_{trap}$) and free carrier density ($n_{band}$) can be obtained by integrating DOS and the Fermi-Dirac distribution over the energy:

$$n_{trap} = \int_{-\infty}^{E_c} DOS_{trap}(E_k) f_{FD} dE_k = \int_{-\infty}^{E_c} \frac{N_T}{k_B T_{ta}} e^{\frac{(E_k - E_c)}{k_B T_{ta}}} f_{FD} dE_k, \quad (3)$$

$$n_{band} = \int_{E_c}^{\infty} DOS_{band}(E_k) f_{FD} dE_k = \int_{E_c}^{\infty} \frac{gm^*}{2\pi\hbar^2} f_{FD} dE_k, \quad (4)$$

$$f_{FD} = \frac{1}{1 + e^{\left(\frac{E_k - E_f}{k_B T}\right)}}. \quad (5)$$

where $f_{FD}$ is the Fermi-Dirac function, $T$ is the temperature, $E_f$ is the fermi level. The ratio of free carrier density ($n_{band}$) over total carrier density ($n_{total}$) is defined as MTR factor ($P_{MTR}$):

$$P_{MTR} = \frac{n_{band}}{n_{band} + n_{trap}} = \frac{n_{band}}{n_{total}} \quad (6)$$

Carriers in the extended states of a-IGZO are subject to various scattering mechanisms. In this work, trapped carrier (TC) scattering and polar optical phonon (PO) scattering are considered. The trapped carrier scattering arises from Coulomb interactions between mobile carriers and charged immobile carriers. The trapped carrier scattering limited mobility $\mu_{TC}$ is given by (2,4):

$$\mu_{TC} = \frac{8\pi\hbar^3(\varepsilon_0\varepsilon_s)^2 k^3 d}{e^3 m^{*2} n_{trap}} \int_0^\pi \frac{\sin^2\theta}{(\sin\theta + \beta)^2} d\theta, \quad (7)$$

where $\beta = S_0/2k$, $S_0$ is the screening constant, $\varepsilon_s$ is static dielectric constant of a-IGZO (~12), $k$ is the wave vector, $\theta$ is the scattering angle and $d$ is effective channel thickness obtained from quantum well approximation. In two-dimensional condition, screening parameter $S_0$ and the wave vector $k$ are expressed depending on condition (5)

$$S_0 = \frac{e^2 n_{band}}{2\varepsilon_0\varepsilon_s k_B T} \text{ and } k = \frac{\sqrt{2k_B T m^*}}{\hbar} \text{ (under non-degenerate condition)} \quad (8)$$

$$S_0 = \frac{2e^2 m^*}{4\pi\varepsilon_s \hbar^2} \text{ and } k = \sqrt{2\pi n_{band}} \text{ (under degenerate condition)} \quad (9)$$

Also, polar optical phonon (OP) limited carrier mobility $\mu_{PO}$ is given by (2,4):

$$\mu_{OP} = \frac{4\pi\hbar^2 \varepsilon_0 \varepsilon_\infty \varepsilon_s}{e\omega d m^{*2}(\varepsilon_s - \varepsilon_\infty)} \left[\exp\left(\frac{\hbar\omega}{k_B T}\right) - 1\right], \quad (10)$$

where $\varepsilon_\infty$ is high frequency dielectric constant of IGZO (~4) (6) and $\omega$ is polar optical phonon frequency.

Therefore, mobility of the carrier in extended states, $\mu_0$ is obtained by using Matthiessen's rule:

$$\frac{1}{\mu_0} = \frac{1}{\mu_{TC}} + \frac{1}{\mu_{OP}} \qquad (11)$$

To describe band transport, carriers in the band states with a path length ($l$) smaller than the lattice constant ($a$) are considered effectively localized (i.e., $l < a$), while only carriers with $l > a$ contribute to band transport. The $l$ of carriers in the extended states are assumed to follow a statistical distribution with a mean value equal to the $l_{mfp}$ (7). And $l_{mfp}$ is defined as (3):

$$l_{mfp} = \begin{cases} \frac{\mu_0}{e}\sqrt{2k_B T m^*} & \text{(non-degenerate)}, \\ \frac{\hbar\mu_0}{e}\sqrt{2\pi n_{band}} & \text{(degenerate)}. \end{cases} \qquad (12)$$

Therefore, not all carriers occupying the extended states contribute to band transport. To account for this, a phenomenological transport reduction factor ($P_{TRF}$) is introduced and defined as the probability that a carrier transverses a distance equal to or greater than lattice constant $a$, based on statistical distribution of path length. (7,8) By assuming an exponential distribution for the free path length with a mean value of $l_{mfp}$, the $P_{TRF}$ is expressed as:

$$P_{TRF} = exp(-a/l_{mfp}). \qquad (13)$$

When $l_{mfp} >> a$ (i.e., $P_{TRF} \sim 1$), it indicates that carriers in the extended states are fully delocalized and transport can be described by band transport. In contrast, when $l_{mfp} << a$ ($P_{TRF} << 1$), it implies that carriers are effectively localized and the band transport is suppressed. By incorporating both the multiple and trap release factor and the transport reduction factor, the overall mobility of the a-IGZO TFT is expressed as

$$\mu_{thin\ film} = P_{MTR}P_{TRF}\mu_{TRF} = P_{MTR}P_{TRF}(1 - lnP_{TRF})\mu_0, \qquad (14)$$

where $\mu_{TRF} = (1 - lnP_{TRF})\mu_0$. Hopping transport is neglected because its contribution to the total current is expected to be small in amorphous oxide semiconductors.

When channel length of device becomes short and when a high lateral electric field is applied along the channel, carriers trapped in localized states can be activated into extended states through both thermal activation and field assisted tunneling (3). Under these high electric field condition, the conventional Fermi-Dirac distribution is not sufficient to describe the carrier population. Instead, a modified Fermi-Dirac function with an effective carrier temperature is employed,

$$f_{FD} = \frac{1}{1+e^{(\frac{E_k-E_f}{k_BT_{eff}})}}. \tag{15}$$

Here, the effective temperature, $T_{eff}$ accounts for field-assisted activation and is defined as

$$T_{eff} = \sqrt{T^2 + (\Upsilon eaE/k_B)^2}, \tag{16}$$

where $E$ is lateral electric field in the channel and $\Upsilon$ is assumed to be 1.

Carriers in extended states gain energy from the lateral electric field and are eventually limited by optical phonon emission, leading to velocity saturation in the band transport. Carrier velocity of amorphous oxide semiconductor is governed not only by the band transport velocity but also the fraction of carriers occupying the extended states. By incorporating field-assisted activation and velocity saturation in the extended states, the ensemble carrier velocity of amorphous oxide semiconductor can be expressed as:

$$v = \frac{P_{MTR}P_{TRF}\mu_{TRF}E}{[1+(\mu_{TRF}E/v_{sat})^\beta]^{\frac{1}{\beta}}}, \tag{17}$$

where $v_{sat} = \sqrt{8\hbar\omega/(3\pi m^*)}$ is the saturation velocity of trap-free IGZO, and $\beta = 2$ for electron transport. In the low-field regime, the velocity reduces to $v \approx P_{MTR}P_{TRF}\mu_{TRF}E$, whereas at high electric fields it saturates at $v_{sat} \approx P_{MTR}P_{TRF}v_{sat}$ rather than $v_{sat0}$.

**Supporting Information 2**

**Modeling a-IGZO TFTs**

In this work, the vertical (perpendicular to the layers; z direction) and lateral (along the channel; y direction) directions are two primary dimensions while channel width direction (x-direction) is assumed to be uniform and homogeneous. The physical thickness of the IGZO channel is 6 nm but primary conduction occurs within a much thinner region near the gate insulator ($Al_2O_3$), defined by quantum confinement width. The vertical dimension is assumed to be a sheet of negligible thickness, and the device can be described using a coupled quasi-2D Poisson equation and a 1D current continuity equation (3, 9):

$$\frac{d^2V(y)}{dy^2} + \frac{C_{ox}(V_{ov}-V(y))}{d\varepsilon_0\varepsilon_s} = \frac{-en_{total}(y)}{d\varepsilon_0\varepsilon_s} \qquad (18)$$

where $V(y)$ is the voltage in the channel, $C_{ox}$ is gate oxide capacitance, $V_{ov}$ is overdrive voltage, $d$ is the effective thickness of channel defined by quantum width. An accurate prediction of threshold voltage $V_T$ as a function of both $V_{GS}$ and $V_{DS}$ is nontrivial. Therefore, $V_T$ is defined as the $V_{GS}$ at which $I_{DS}$ reaches 10 nAμm$^{-1}$ at $V_{DS}$ = 1 V, and it is assumed to remain approximately constant over the drain bias range. A more rigorous $V_T(V_{GS}, V_{DS})$ model can be incorporated into the same self-consistent framework in future work. Under steady state, the drain current $I_d$ is constant across the channel and is given by:

$$I_d = Wen_{total}(y)v(y), \qquad (19)$$

where $W$ is channel width, $n_s(y)$ is total carrier density in the channel and $v(y)$ is the carrier velocity in the channel as described by **Eq. 3** in main manuscript. In this work, the I-V characteristics of the a-IGZO TFTs are obtained through a self-consistent solution of the coupled quasi-2D Poisson equation and current continuity equation, as described in **Fig. S1**.

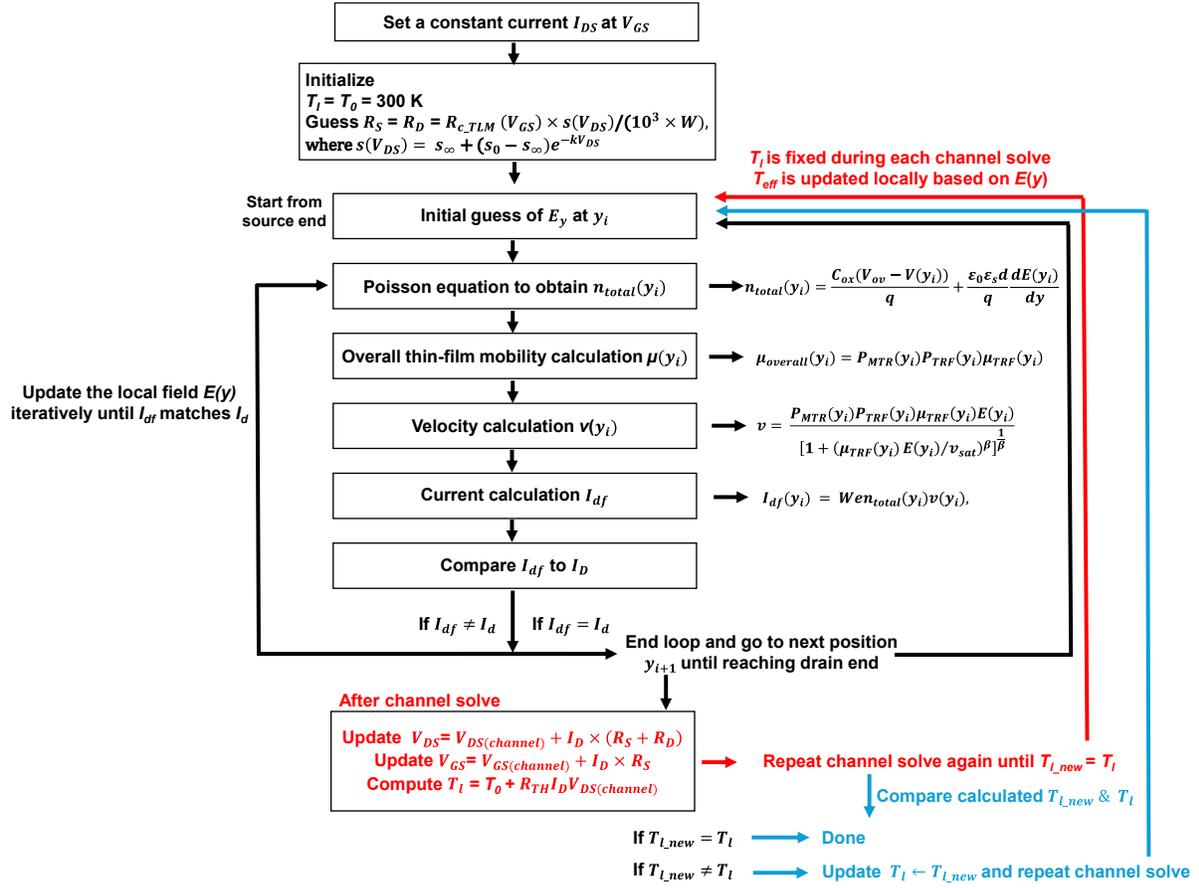

**Figure S1**. Self-consistent simulation flowchart for the quasi-2D TFT transport model.

**Figure S1** shows simulation flowchart used in this study. The coupled quasi-2D Poisson equation and the current continuity equation are solved using a self-consistent iterative scheme. For a given external gate and drain bias, the solution proceeds as follows:

1) The channel is discretized from the source ($y = 0$) to the drain ($y = L_{ch}$).
2) For a given external gate voltage $V_{GS}$, a constant drain current $I_{DS}$ is assumed along the channel.

3) The lattice temperature ($T_l$) is initialized to the ambient temperature ($T_l = T_0 = 300$ K). The source and drain series resistance ($R_S$ and $R_D$) are initialized using the contact resistance extracted from the transmission line method (TLM) and $V_{DS}$ dependent contact resistance factor.

$$R_S = R_D = \frac{R_{c\_TLM}(V_{GS}) \times s(V_{DS})}{(10^3 \times W)}, \qquad (20)$$

where $s(V_{DS})$ accounts for drain bias dependence of the contact resistance.

4) An initial value of the lateral electric field ($E_y$) is set at the source end ($y=0$).

5) At each position, $y_i$, the channel voltage $V(y_i)$ is obtained by integrating the lateral electric field and the total carrier density $n_{\text{total}}(y_i)$ is calculated using the quasi-2D Poisson equation.

6) Overall thin film mobility is calculated as:

$$\mu_{overall}(y_i) = P_{MTR}(y_i)P_{TRF}(y_i)\mu_{TRF}(y_i) \qquad (21)$$

7) The effective carrier velocity is calculated as:

$$v(y_i) = \frac{P_{MTR}(y_i)P_{TRF}(y_i)\mu_{TRF}(y_i)E(y_i)}{[1+(\mu_{TRF}(y_i)E(y_i)/v_{sat})^\beta]^{\frac{1}{\beta}}} \qquad (22)$$

8) The local current $I_{df}$ is calculated as follows:

$$I_{df}(y_i) = Wqn_{total}(y_i)v(y_i). \qquad (23)$$

The calculated $I_{df}(y_i)$ is compared with previously assumed $I_{DS}$ in step 2. If $I_{df}(y_i) \neq I_{DS}$, a new value of $E_y$ is chosen and above steps are repeated. If $I_{df}(y_i) = I_{DS}$, the $V(y_i)$, $E(y_i)$, $n_{total}(y_i)$ etc are obtained.

9) The next position $y_{i+1}$ is then processed and above steps are repeated sequentially along the channel until the drain end is reached.

10) After completing above steps, the external voltages are updated self-consistently as:

$$V_{DS} = V_{DS(channel)} + I_D \times (R_S + R_D), \qquad (24)$$

$$V_{GS} = V_{GS(channel)} + I_D \times R_S. \quad (25)$$

And the lattice temperature is recalculated using the following equation

$$T_l = T_0 + R_{TH}I_D V_{DS(channel)}. \quad (26)$$

11) Solving channel is repeated with the updated $T_l$ until the lattice temperature converges.

The material, transport, and thermal parameters used in the modeling are summarized in **Table S1**.

**Table S1**. Parameters of a-IGZO TFT simulation

| | | | |
|---|---|---|---|
| $W_{ch}$ | 3.4 μm | $\varepsilon_s$ | 12 |
| $L_{ch}$ | 100 nm, 50 nm | $\varepsilon_\infty$ | 4 |
| $t_{ox}$ | 9 nm | $\varepsilon_{ox}$ | 7.5 |
| $\beta$ | 2 | $\hbar w_0$ | 20 meV |
| $m^*$ | 0.34 $m_0$ | $a$ | 0.33 nm |
| $N_t$ | 6.5×10$^{12}$ cm$^{-2}$ | $Tta$ | 1150 K |
| $d$ | Quantum-confined width | | |

**Thermal modeling**

Self-heating in the a-IGZO TFTs is modeled using a compact thermal resistance ($R_{TH}$) framework that captures the dominant heat dissipation pathway. Instead of solving the full spatial temperature distribution, the lattice temperature rise ($\Delta T_l$) is expressed as

$$\Delta T_l \approx R_{TH}I_D V_{channel} \quad (27)$$

where $V_{channel}$ is channel voltage obtained from self-consistent calculation. The generated heat in the channel is assumed to dissipate primarily in the vertical direction through the gate stack and the substrate. Using the channel area ($A = L_{ch} \times W_{ch}$), the vertical thermal resistance ($R_{TH}$) is modeled as a series combination of the IGZO channel thermal resistance ($R_{channel}$), the IGZO/Al$_2$O$_3$ thermal boundary resistance ($R_{TBR}$), the $Al_2O_3$ dielectric layer thermal resistance ($R_{oxide}$), the $Al_2O_3$/Ni thermal boundary resistance ($R_{TBR}$) and Ni contacts thermal resistance ($R_{contact}$):

$$R_{TH} = R_{channel} + R_{TBR}^{IGZO/Al_2O_3} + R_{oxide} + R_{TBR}^{Al_2O_3/Ni} + R_{contact}. \quad (28)$$

The thermal conductivity ($k$) and thermal boundary conductance ($G$) values used in the calculation are summarized in **Table S2**. This estimation yields $R_{TH} \sim 1.1 \times 10^5$ K/W for purely vertical heat flow confined to channel area. In practice, lateral heat spreading in the gate stack can increase the effective heat-flow area and reduce the effective thermal resistance. Accordingly, an effective thermal resistance value of $R_{TH} = 9 \times 10^4$ K/W is used in the study. This lattice temperature modifies phonon-limited mobility and effective carrier temperature, thereby affecting carrier velocity and the drain current.

**Table S2**. Thermal properties of a-IGZO TFT simulation

| Material | $k$ (Wm$^{-1}$K$^{-1}$) | Material Interface | $G$ (MWm$^{-2}$K$^{-1}$) |
|---|---|---|---|
| IGZO (10,11) | 1.5 | IGZO–$Al_2O_3$* | 50 |
| $Al_2O_3$ (12,13) | 1.5 | $Al_2O_3$–Ni (12) | 150 |
| Ni (12) | 40 | | |

* The thermal boundary conductance (G) value of IGZO–$Al_2O_3$ is approximated by $G$ values for ITO–$Al_2O_3$ (12).

**Supporting Information 3**

**Thin film mobility and measured low-field mobility**

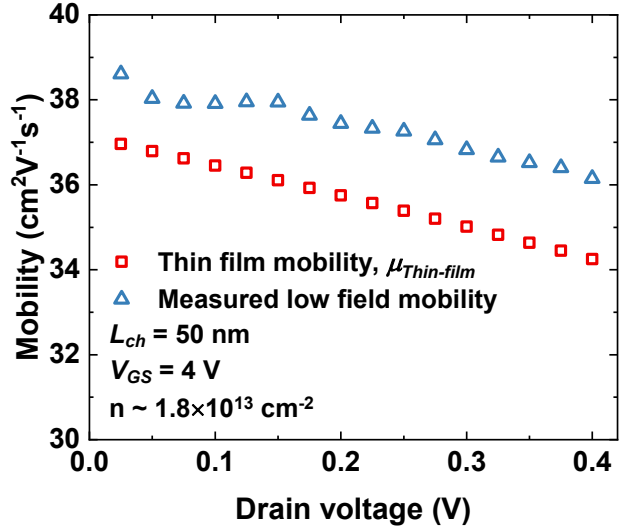

**Figure S2**. Comparison between thin film mobility and experimentally measured low-field mobility of 50 nm $L_{ch}$ device at $n \sim 1.8 \times 10^{13}$ cm$^{-2}$.

Thin film mobility, $\mu_{Thin-film}$ is calculated using Equation 14 above.

$$\mu_{Thin-film} = P_{MTR}P_{TRF}\mu_{TRF} = P_{MTR}P_{TRF}(1 - \ln P_{TRF})\mu_0.$$

The measured low-field mobility is extracted from the experimental data as

$$\mu = \frac{I_D}{\frac{W}{L}C_{ox}(V_{GS\_channel} - V_T)V_{DS\_channel}}, \quad (29)$$

where

$$V_{GS\_channel} = V_{GS} - I_D \times R_S \quad (30)$$

and

$$V_{DS\_channel} = V_{DS} - I_D \times (R_S + R_D). \quad (31)$$

Using two method, mobility ~38 cm$^2$V$^{-1}$s$^{-1}$ at $n \sim 1.8 \times 10^{13}$ cm$^{-2}$ is obtained. Several studies have also reported high mobility at high carrier density (14,15)

**Supporting Information 4**

**TLM based contact resistance extraction**

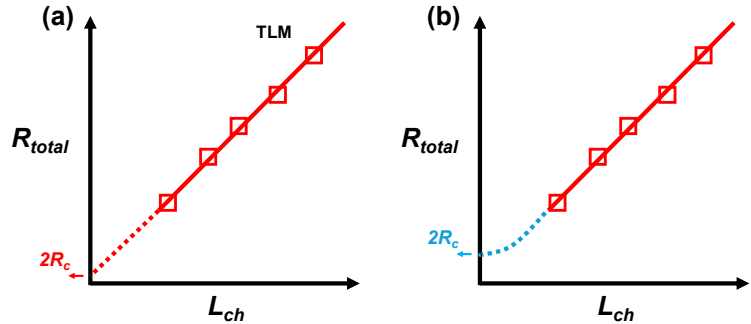

**Figure S3**. Schematic illustration of $2R_c$ extraction using the conventional TLM method, where a linear $R_{total}$ vs. $L_{ch}$ relation is assumed down to $L_{ch} = 0$. (b) Possible scenario in short-channel devices, where non-linearity in the $R_{total}$ vs. $L_{ch}$ relation near small channel lengths leads to a larger intercept and thus a higher extracted $2R_c$ compared to the TLM method.

**Supporting Information 5**

**50 nm $L_{ch}$ device calibration data**

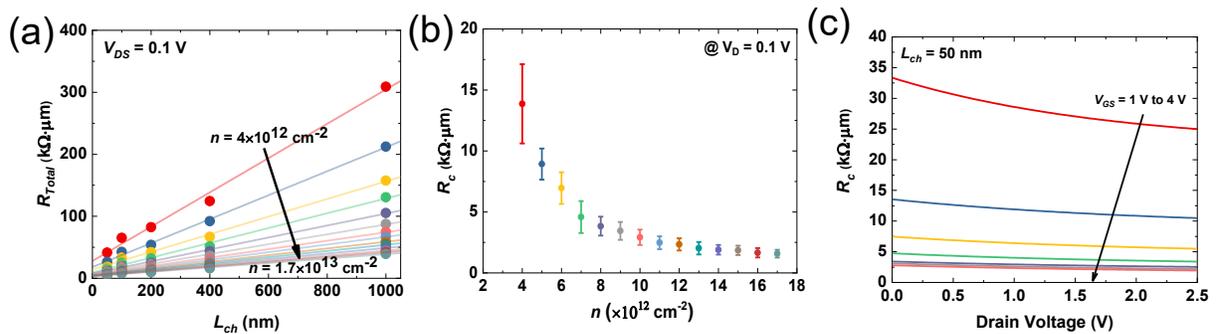

**Figure S4**. (a) Total resistance ($R_{total}$) as function of channel length measured at $V_{DS} = 0.1$ V for different carrier densities. Symbols represent experimental data and solid lines are linear fits. (b) Extracted $R_c$ obtained from the y-intercepts of the linear fits to $R_{total}$ vs. $L_{ch}$ in **Fig. S3a**. Error bars represent the standard errors of the fitted intercepts. (c) $V_G$ and $V_D$ dependent $R_c$ used in the model.

**Supporting Information 6**

**Extracted spatial profiles along the channel of a 50 nm $L_{ch}$ device**

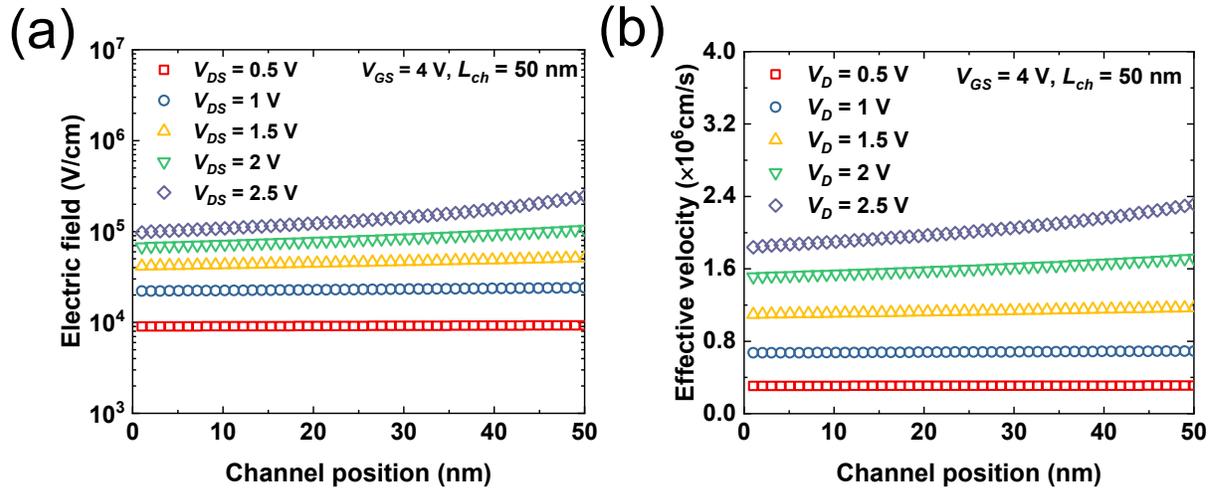

**Figure S5**. Extracted spatial profiles along the channel of a 50 nm $L_{ch}$ device at $V_{GS}$ = 4 V: (a) Lateral local electric field and (b) effective carrier velocity.